\documentclass{article}
\usepackage{spconfa4,amsmath,graphicx}
\usepackage{cite}
\usepackage{amsmath,amssymb,amsfonts}
\usepackage{graphicx}
\usepackage{textcomp}
\usepackage{xcolor}

\usepackage{hyperref}

\usepackage{algpseudocode}

\usepackage[normalem]{ulem}
\useunder{\uline}{\ul}{}
\usepackage{graphicx}
\usepackage{subcaption}
\usepackage{mwe}
\usepackage{comment}
\usepackage{amsmath}
\usepackage{algpseudocode}
\usepackage{algorithm}
\usepackage{lipsum}
\usepackage[export]{adjustbox}
\usepackage{amsthm}

\algnewcommand\algorithmicforeach{\textbf{for each}}
\algdef{S}[FOR]{ForEach}[1]{\algorithmicforeach\ #1\ \algorithmicdo}

\def\x{{\mathbf x}}
\def\L{{\cal L}}

\def\by{{\mathbf y}}

\def\E{{\mathbb E}}
\def\L{{\mathcal L}}
\def\GP{{\mathcal G \mathcal P}}
\def\R{{\mathbb R}}
\def\cN{{\mathcal N}}

\def\hbu{{\hat{\mathbf{u}}}}
\def\X{\mathbf{X}}
\def\bTheta{\mathbf{\Theta}}
\def\bb{\mathbf{b}}

\def\K{{\mathbf{K}_{\X,\X}^{u}}}
\def\hK{{\hat{\mathbf{K}}_{\X,\X}^{u}}}
\def\Kpred{{\mathbf{K}_{\hat{\X},\X}^{u}}}
\def\KpredT{{\mathbf{K}^{u}_{\X,\hat{\X}}}}
\def\Kpredpred{{\mathbf{K}_{\hat{\X},\hat{\X}}^{u}}}

\def\Kxz{{\mathbf{K}_{\X,\X_z}^{u,z}}}
\def\Kzx{{\mathbf{K}_{\X_z,\X}^{z,u}}}
\def\Kzz{{\mathbf{K}_{\X_z,\X_z}^{z,z}}}
\def\hKzz{{\hat{\mathbf{K}}_{\X_z,\X_z}^{z,z}}}

\def\tK{{\Tilde{\mathbf{K}}_{\X,\X}}}

\def\hbr{\hat{\mathbf{r}}}

\def\bY{{\mathbf Y}}
\def\cL{{\mathcal L}}
\def\tbY{\Tilde{\mathbf{Y}}}
\def\bZ{\mathbf{Z}}

\usepackage{amsmath,amssymb,amsfonts}

\newcommand{\br}{\mathbf{r}}

\def\minwrt[#1]{\underset{#1}{\mathrm{minimize }}}
\def\argminwrt[#1]{\underset{#1}{\mathrm{arg min }}}

\floatstyle{ruled}
\newfloat{algorithm}{t}{lop}
\floatname{algorithm}{Algorithm}

\newcommand{\bu}{\mathbf{u}}

\begin{document}

\title{Sound Field Estimation Using Deep Kernel Learning \\ Regularized by the Wave Equation}

\name{David Sundström$^{\star}$ \qquad Shoichi Koyama $^{\dagger}$ \quad Andreas Jakobsson$^{\star}$}

\address{$^{\star}$ Dept. of Mathematical Sciences, Lund University, Sweden \\
$^{\dagger}$National Institute of Informatics, Japan}

\maketitle
\begin{abstract}
In this work, we introduce a spatio-temporal kernel for Gaussian process (GP) regression-based sound field estimation. Notably, GPs have the attractive property that the sound field is a linear function of the measurements, allowing the field to be estimated efficiently from distributed microphone measurements. However, to ensure analytical tractability, most existing kernels for sound field estimation have been formulated in the frequency domain, formed independently for each frequency. 
To address the analytical intractability of spatio-temporal kernels, we here propose to instead learn the kernel directly from data by the means of deep kernel learning.
Furthermore, to improve the generalization of the deep kernel, we propose a method for regularizing the learning process using the wave equation. The representational advantages of the deep kernel and the improved generalization obtained by using the wave equation regularization are illustrated using numerical simulations.
\end{abstract}


\section{Introduction}

Accurate sound field estimation is essential for spatial audio applications such as spatial active noise control \cite{Koyama2021}, the generation of individual sound zones \cite{Lee2018}, and virtual/augmented reality \cite{poletti2005three}. Given a stream of signals from a microphone array, the problem is to estimate the sound field in a region of interest. Due to requirements for fast updates of the sound field estimate when new signals arrive, kernel methods, i.e., models of the spatial covariance function, have been popular for sound field estimation, resulting in the estimation to be a linear mapping from the measurements \cite{Caviedes2021gaussian}. However, it remains a challenging problem to accurately represent the kernel for a sound field generated by broadband sources within a room. In particular, the impulse response from a source to a receiver consists of both temporal and spectral correlations, resulting in a rich spatio-temporal structure of the observed signals. Due to the inherent complexity of the sound field, the formulation of an analytical kernel function in the time domain remains notoriously difficult.

There is a rich literature on kernel functions that satisfy the Helmholtz equation in the frequency domain, including diffuse kernels \cite{Ueno2018kernel}, directionally weighted kernels \cite{ueno2021directionally,sundstrom2023recursive}, and kernels parameterized by neural networks \cite{ribeiro2023sound}. However, for the frequency-domain kernels, it is common to make the approximation that different frequency components act independently in order to allow for analytically tractable formulations. 
While the literature considering time-domain kernel formulations is more limited, other approaches for representing the spatio-temporal sound field include the equivalent source model \cite{antonello2017room} as well as deep parameterizations \cite{olivieri2024physics,lluis2020sound}, which requires solving large optimization problems when new data is observed. 

In this work, we introduce a method for learning a spatio-temporal kernel directly from the microphone measurements by means of deep kernel learning \cite{Wilson2016deep}.
To the best of the authors' knowledge, this is the first time a spatio-temporal kernel formulation is proposed for sound field estimation. Furthermore, to improve the generalization of the model to unseen positions, we propose to regularize the kernel estimation by incorporating the homogeneous wave equation in a similar manner to the physics-informed neural network~\cite{Raissi2019}. The proposed physics-informed deep kernel learning approach is valid for any known linear PDE.

\section{Preliminaries}
\label{sect:background}
\subsection{Problem formulation}
\label{sect:problem_formulation}
Consider a sound field, $u: \R^4 \rightarrow \R$, represented as a function of a position $\br\in \R^3$ and time $t\in \R$, satisfying the wave equation 
\begin{equation}
    \L_\x u(\x) = z(\x),
    \label{eq:wave_equation}
\end{equation}
where $\x = [\br \quad t]^T$ and the operator $\L_\x$ being defined as
\begin{equation}
\L_\x = \Delta_\br -\frac{1}{c^2}\frac{\partial^2}{\partial t^2},
\end{equation}
with $c$ denoting the speed of sound, $\Delta_\br$ the Laplace operator, and $z:\R^4\rightarrow\R$  the source distribution. For a source-free spatio-temporal domain $\Omega \subseteq \R^4$, as considered in this work,  the source distribution is $z(\x) = 0$, $\forall \x \in \Omega$.
Let $y(\x_d)$ denote a noisy measurement of such a sound field, such that
\begin{equation}
    y(\x_d) = u(\x_d)+\epsilon_{d},
    \label{eq:signal_model}
\end{equation}
where $\epsilon_{d}$ is an additive noise term and $\x_d = [\br_d \quad t_d]^T$. Furthermore, let $\by \in \R^{NM}$ be a vector of $NM$ measurements of the sound field measured at $M$ microphone positions and $N$ time points, with  
\begin{equation}
\by = \left[ \begin{array}{ccc} y(\x_1) & \ldots & y(\x_{NM}) \end{array} \right]^T. 
\end{equation}
Given the measurements $\by$, we here consider the problem of reconstructing the sound field at another position and time $[\hbr \quad \hat{t}]^T\in \Omega$. Due to the importance of being able to form computationally efficient updates of the sound field as new data is available, we focus our attention to processes that may be well approximated as GPs.

\subsection{Gaussian processes for sound field estimation}
We proceed by introducing the framework of GPs for sound field modeling (see, e.g., \cite{Rasmussen2003gaussian} for an overview of GPs). 
To do so, the sound field, $u$, is modelled as a zero-mean GP, $u \sim \GP(0,\kappa_u)$, where $\kappa_u: \R^{4 \times 4} \rightarrow \R$ is a positive definite kernel. The GP prior over $u$ is defined such that point-wise evaluations, $\bu = [u(\x_1) \quad \hdots \quad u(\x_{NM})]^T$, follow a multi-variate normal distribution $\bu\sim \cN(0, \K)$, where 
\begin{equation}
\X = \left[ \begin{array}{ccc} \x_1 & \ldots & \x_{NM} \end{array} \right]^T 
\in \R^{NM \times 4},
\end{equation}
denotes the measurement grid, and 
the Gram matrix $\K\in \R^{NM\times NM}$ is formed as $[\K]_{i,j} = \kappa_u(\x_i,\x_j)$.  
Assuming that the observation noise in \eqref{eq:signal_model} may be well modeled as being white and Gaussian distributed, i.e., $\epsilon_{d}\propto \cN(0,\sigma^2)$, 
then both the measurements, $\bY$, and the predictive distribution, $\hbu$, may be detailed as
\begin{align}
    \bY &\sim \cN(0,\K+\sigma^2I), \\
    \hbu &\sim \cN(\E(\hbu), cov(\hbu)),
\end{align}
where the expectation and covariance of the predictive distribution are given by
\begin{align}
    \E(\hbu) &= \Kpred(\K+\sigma^2I)^{-1}\bY, \\
    cov(\hbu) &= \Kpredpred - \Kpred (\K+\sigma^2 I)^{-1}\KpredT,
\end{align}
with $\hat{\X}$ denoting the $P$ prediction positions and the predictive covariance matrices, $\Kpred\in \R^{P \times NM}$ and $\Kpredpred\in \R^{P \times P}$, are defined similarly to $\K$. 
Consequently, by modeling the sound field as a GP, the expectation of the predictive distribution is a linear mapping with respect to the measurements. This is an attractive property for applications within audio signal processing, since it allows for fast predictions when new observations, possibly both in time and space, are available. 
However, the accuracy of the predictions relies on the function space of the GP, defined by the kernel function $\kappa_u$. While several analytical kernel functions have been proposed in the frequency domain, the analytic formulation of time-domain kernel functions for sound field estimation remains a challenging problem. Here, we approach this problem by approximating a time-domain kernel function from the available data by means of deep kernel learning.

\subsection{Deep kernel learning}
The concept of exploiting the expressiveness of deep neural networks to model involved kernel functions was recently presented in \cite{Wilson2016deep} under the name deep kernel learning. The idea is to warp the input space of the kernel by a non-linear mapping, i.e., $\phi:\R^{4} \rightarrow \R^{h}$, where $\phi$ is parameterized by a deep neural network, with $h$ being the dimension of the feature representation. The deep kernel, $\kappa_{\mathrm{DK}}$, is then formulated on the form
\begin{equation}
    \kappa_{\mathrm{DK}}(\x_i, \x_j) = \kappa_u(\phi(\x_i),\phi(\x_j)).
    \label{eq:deep_kernel}
\end{equation}
The foundation for \eqref{eq:deep_kernel} being a valid kernel, given that $\kappa_u$ is a valid kernel, was studied in \cite{Bohn2019}. Without loss of generality, we assume the kernel function %$\kappa_u$ to be a squared exponential kernel, such that 
\begin{equation}
    \kappa_u(\x_i,\x_j) = \sigma_\kappa^2 e^{-||\x_j-\x_i||_2^2/2\ell^2},
    \label{eq:squared_exponential_kernel}
\end{equation}
where $\sigma_\kappa \in \R$ denotes the standard deviation and $\ell\in\R$ the length-scale. The squared exponential kernel in \eqref{eq:squared_exponential_kernel} is a universal kernel and is infinitely differentiable \cite{Micchelli2006universal}. The parameters $\theta = \{\gamma, \ell, \sigma_\kappa  \}$ of the deep kernel in \eqref{eq:deep_kernel} does thus both contain the parameters of the neural network, $\gamma$, and the parameters of the kernel, i.e., $\ell$ and $\sigma_\kappa$. In \cite{Wilson2016deep}, the parameters $\theta$ were estimated by maximizing the marginal log-likelihood, i.e., the probability of the data conditioned on the hyperparameters, which for a GP is given by
\begin{align}
 %   \mathcal{L}_{\bY} &=
    \log p(\bY | \theta,\X) %\nonumber \\
    &
    = -  \bY^T(\K+\sigma^2I)^{-1}\bY \nonumber \\
    & \hspace{12mm}
    - \log|\K + \sigma^2I | + \mbox{const}.
    \label{eq:marginal_likelihood}
\end{align}
The log-likelihood includes two (non-constant) terms, one which adapts the kernel to the data and one which naturally regularizes the complexity of the kernel. Despite the self-regularization property, the issue of {\it feature collapse} has been observed when learning a deep kernel using the marginal log-likelihood \cite{vanAmersfoort2021,Ober2021,Calandra2016}. 
In \cite{Ober2021}, it is argued that the large parameter space causes the learning to correlate all data-points, rather than just those for which correlations are important. Various ways to get around the feature collapse have been proposed, including imposing bi-Lipschitz constraints on $\phi$ \cite{vanAmersfoort2021,Liu2020}, fully Bayesian approaches \cite{Ober2021}, and pre-training of the feature space \cite{Chen2022}. For the sound field estimation problem considered in this work, we instead proceed to restrict the parameter space by introducing wave equation priors on the kernel function.

\section{Deep kernel learning regularized \\ by the wave equation}
\label{sect:method}
\vspace{-2mm}
In this section, we introduce a method for using the homogeneous wave equation as prior when estimating the parameters of the deep kernel in \eqref{eq:deep_kernel} using the marginal log-likelihood in \eqref{eq:marginal_likelihood}. 
Reminiscent to the work in \cite{Raissi2017machine}, which proposes to use the joint distribution of $u$ and $z$ for the shallow kernel setting, we show the advantage of incorporating priors on $z$ in the deep kernel learning setting and propose to exploit the block-structure to improve the conditioning of the problem.

Proceeding, it may be noted that the operator $\cL_{\x}$ in  \eqref{eq:wave_equation} is linear, implying that $z$ is a zero-mean GP, i.e., $z\sim \mathcal{GP}(0,\kappa_z)$, with the covariance function 
\begin{equation}
    \kappa_z(\x_i,\x_j) = \L_{\x_i} \L_{\x_j} \kappa_u(\x_i,\x_j),
    \label{eq:k_z}
\end{equation}
following from that 
\begin{align}
    \E \{\L_{\x_i}u(\x_i)\L_{\x_j}u(\x_j) \} = \L_{\x_i} \L_{\x_j}  \E \{u(\x_i)u(\x_j) \},
\end{align}
for any linear PDE \cite{Raissi2017machine}. Thus, the parameters $\theta$ parameterizes both the kernel of the GP $u$ and $z$. We here propose to estimate these parameters by minimizing the negative log marginal likelihood of the joint distribution of $u$ and $z$, i.e., 
\begin{equation}
    \min_{\theta} \tbY^T\tK^{-1}\tbY + \log |\tK|,
    \label{eq:problem}
\end{equation}
where $\tbY = [\bY \quad \bZ]^T \in \R^{NM + N_zM_z}$, $\bZ \in \R^{N_zM_z}$ are the observations of $z$ at $M_z$ positions in space and $N_z$ samples in time. It is worth noting that the collocation points for which $z$ is evaluated do not necessarily have to coincide with the microphone positions. The $(NM + N_zM_z) \times (NM + N_zM_z)$ Gram matrix, $\tK$, of the joint process is formed as
\begin{equation}
    \tK = 
    \begin{bmatrix}
        \K + \sigma^2I & \Kxz \\
        \Kzx & \Kzz + \sigma_z^2I
    \end{bmatrix},
    \label{eq:K_tilde}
\end{equation}
where $\sigma_z \in \R$ is included to capture noise in the source distribution, the cross-correlation blocks $\Kxz \in \R^{NM \times N_zM_z}$ and $\Kzx \in \R^{N_zM_z \times NM}$ are 
\begin{align}
    [\Kxz]_{i,j} &= \L_{\x_j}\kappa_u(\x_i,\x_j), \\
    [\Kzx]_{i,j} &= \L_{\x_i}\kappa_u(\x_i,\x_j),
\end{align}
respectively, and the $N_zM_z \times N_zM_z$ Gram matrix of the source distribution, $\Kzz$, is defined as 
\begin{align}
[\Kzz]_{\x_i,\x_j} = \L_{\x_i}\L_{\x_j}\kappa_u(\x_i,\x_j).
\end{align}
While the formulation in \eqref{eq:problem} allows for both measurements of the sound field, $\bY$, as well as the source distribution, $\bZ$, only the sound field is typically measured for spatial audio applications. Instead, we use the vector $\bZ$ to incorporate the assumption of a homogeneous sound field at the $N_zM_z$ collocation points by setting it to the zero-vector $\mathbf{0}\in \R^{N_zM_z}$. 

Finally, we note that the $(NM+N_zM_z)\times (NM+N_zM_z)$ matrix in \eqref{eq:K_tilde} grows rapidly with increasing number of measurements and collocations points. In addition, since the kernels of each block in \eqref{eq:K_tilde} are of varying magnitudes, the computation of the determinant and inverse in \eqref{eq:problem} may be ill-conditioned due to the large eigenvalue-spread. To improve the conditioning, we exploit the block-structure when computing the inverse and determinant in \eqref{eq:problem}, and equivalently state the problem as  \cite{petersen2008matrix}
\begin{equation}
\begin{aligned}
    \min_\theta & \bY^T (\hK -\Kxz(\hKzz)^{-1}\Kzx)^{-1}\bY + \\
    & \hspace{-3mm} \log|\hK||\hKzz-\Kzx(\hK)^{-1}\Kxz|,
\end{aligned}
\label{eq:DKPDE}
\end{equation}
where we have used that $\hK = \K+\sigma^2I$, $\hKzz = \Kzz + \sigma_z^2 I$, and that $\bZ = \mathbf{0}$. As a result, the formulation only requires computing the inverses and determinants of matrices of dimensions $NM\times NM$ and $N_ZM_z\times N_zM_z$. In practice, it is sufficient to use only a few collocation points since new points can be sampled within $\Omega$ at every gradient step when solving \eqref{eq:problem} to achieve a solution which approaches a solution of the wave equation in the %full
domain $\Omega$. 

\begin{figure}
    \centering
    \includegraphics[width = \linewidth]{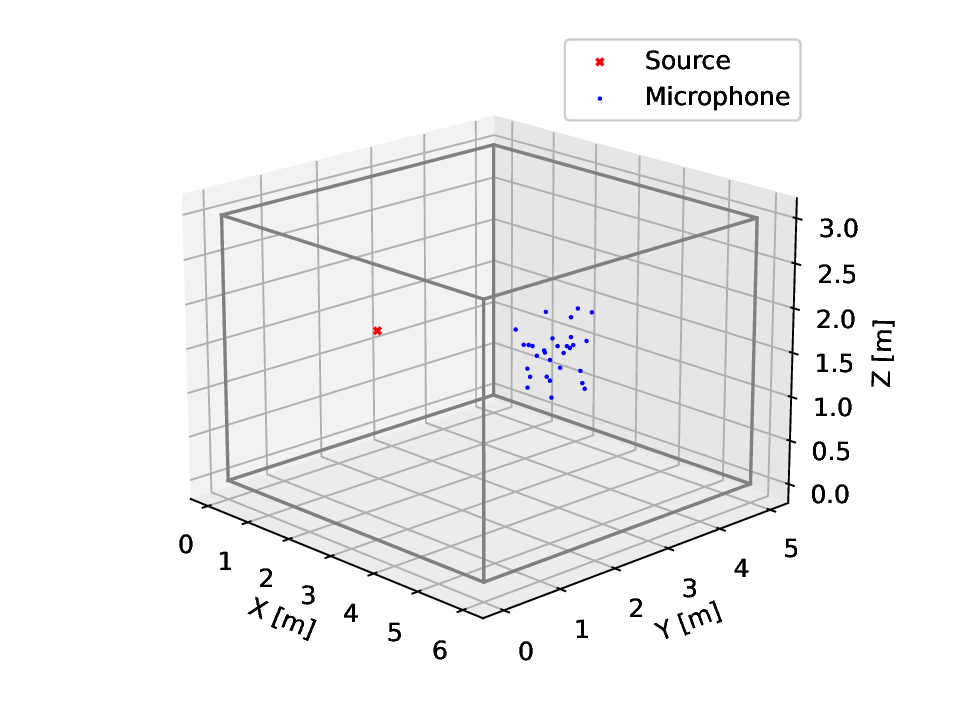}
    \caption{Illustration of the room geometry, microphone positions, and source position for the simulation setup.}
    \label{fig:geometry}
\end{figure}

\begin{figure}
    \centering
    \includegraphics[width = \linewidth]{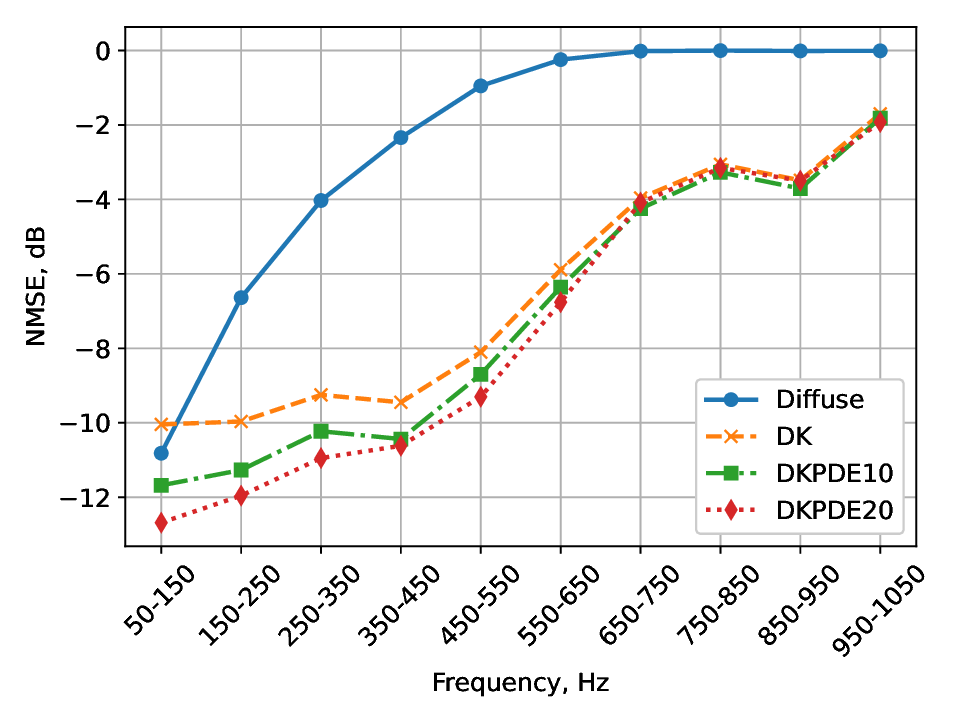}
    \caption{Illustration of the interpolation results when learning the kernel from a broadband source.}
    \label{fig:frequency}
    \vspace{-5mm}
\end{figure}

\section{Numerical experiments}
\label{sect:numerical_experiments}
\vspace{-1mm}
In this section, we illustrate the representational benefits of the deep parameterization when estimating the sound field generated by a broadband source in a room\footnote{The code for the numerical experiments can be found at  \url{https://github.com/davstrom99/dkpde.git}}. We construct a simulation where the kernel is estimated based on a sequence of signal batches of length $50$ samples at the sampling rate of $3$~kHz using an array consisting of $30$ microphones that were randomly placed (uniformly) within a cube of side-length $0.5$~m with a signal to noise ratio of $40$~dB. The microphone array is centered at $[4.0, 3.0, 1.5]~\mathrm{m}$ in a room of dimensions  $6.0\times 5.0 \times 3.0$~m as illustrated in Figure \ref{fig:geometry}, with the sound field being simulated by filtering with room impulse responses generated by the image source method \cite{Allen1979} 
with absorption coefficient $0.55$, maximum order of $26$, and reverberation time $T_{60} = 0.2$~s.
The source signal is white Gaussian noise bandpass filtered to the range $50$~Hz to $1$~kHz and positioned at $[1.0, 2.0, 1.5]$~m.
The proposed method is validated against the kernel method proposed in \cite{Ueno2018kernel}, here termed the diffuse kernel method. To adapt it to the time-domain setting, we construct the diffuse kernel on a grid of $1025$ equally spaced frequency components in the range $50$ to $1$~kHz. 
Furthermore, we present the results of the proposed method detailed above both with and without the wave equation regularization. The deep neural network architecture is described in Appendix~\ref{sect:appendix}. The parameters $\theta$ of the deep kernels are estimated using the Adam optimizer \cite{kingma2014adam} during 20000 epochs with a learning rate of $10^{-5}$. For the wave equation regularization in \eqref{eq:DKPDE}, we randomize new positions for the collocation positions for the constraints at every gradient step within the same region as the microphone array. We present the results for two choices, both 10 and 20, of the
number of collocation points at each gradient step, and denote the corresponding methods DKPDE10 and DKPDE20, respectively, and denote the method using deep kernel learning without wave equation regularization DK. 
In order to evaluate the methods, we use the normalized mean squared error (NMSE), defined as
$\text{NMSE} = {\|u(\x)-\hat{u}(\x)\|_2^2}/{\|u(\x)\|_2^2}$,
%\begin{equation}
%    \text{NMSE} = \frac{\|u(\x)-\hat{u}(\x)\|_2^2}{\|u(\x)\|_2^2},
%    \label{eq:NMSE}
%\end{equation}
where $u$ and $\hat{u}$ denote the true and estimated sound fields, respectively. The NMSE is averaged over the 50 samples long signals at 100 randomly chosen microphone positions in space, for 10 different realizations of the source signal and array geometries.
In Figure \ref{fig:frequency}, the interpolation results are illustrated in different frequency bands, i.e., where $u$ and $\hat{u}$ %in \eqref{eq:NMSE} are bandpass 
filtered to the corresponding interval. As may be seen in the figure, the diffuse method struggles to represent the three-dimensional spatio-temporal wave field. Also, one may note that the PDE regularization consistently improves the interpolation accuracy as more collocation points are added. However, due to the large memory allocation required for computing the gradients of the kernel in \eqref{eq:k_z}, it remains for future work to solve for PDE regularization with a large number of collocation points. The NMSE of the entire signals is shown in Table \ref{tab:NMSE}, and we note that the high NMSE of the diffuse kernel, compared to, e.g., \cite{olivieri2024physics}, reflects the challenging experimental setting.

\begin{table}[t]
\begin{tabular}{l|l|l|l|l|}
\cline{2-5}
\textbf{}                               & \textbf{Diffuse} & \textbf{DK} & \textbf{DKPDE10} & \textbf{DKPDE20} \\ \hline
\multicolumn{1}{|l|}{\textbf{NMSE, dB}} & -1.31            & -6.09       & -6.52            & -6.53            \\ \hline
\end{tabular}
\caption{Interpolation results on the full frequency range from 50 to 1000 Hz.}
\label{tab:NMSE} \vspace{-4mm}
\end{table}

\section{Conclusion}
\label{sect:conclusion}
\vspace{-1mm}
In this work, we propose to represent a spatio-temporal kernel for sound field estimation by means of deep kernel learning. To improve the estimation accuracy at unseen positions, we propose to use the wave equation for regularizing the estimation of the deep kernel. We illustrate the representational benefits of the proposed method using numerical simulations.

%\section{Acknowledgement}
\footnotetext{
The authors would like to acknowledge  Juliano G. C. Ribeiro at the University of Tokyo and Filip Tronarp at Lund University for insightful discussions on Gaussian processes. This work was partly supported by JST FOREST Program, Grant Number JPMJFR216M, and the European Union's ERASMUS+ program.}

\appendix
\section{Neural Network Architecture}
\label{sect:appendix}
\vspace{-1mm}
Since the main contribution of this work is the framework for estimating the proposed deep kernel, we use a simple network architecture which recently has been widely used in the context of sound field estimation \cite{karakonstantis2024room,olivieri2024physics,sitzmann2020implicit}. We summarize the architecture here and refer to \cite{karakonstantis2024room} for details. We use a multiple-layer perceptron structure defined as 
\begin{equation}
    \phi(\x) = \xi_L \circ \xi_{L-1} \circ \hdots \circ \xi_{1}(\x),
\end{equation}
where $\xi_l$ denotes each layer, which are defined as 
\begin{align}
    \xi_l(\x) &= \sin(\omega_0 \x^T \bTheta_l+ \bb_l) \quad \text{for} \quad l=1,\hdots,L-1, \\
    \xi_L(\x) &= \omega_0 \x^T \bTheta_l+ \bb_l,
\end{align}
where $\bb\in\R^H$ is the bias,  $\bTheta_1 \in \R^{H\times 4}$,  $\bTheta_l \in \R^{H\times H}$, for $l=2,\ldots, L$, are the weights, and $\omega_0\in \R$ is a constant controlling the initialization of the frequency. In this work, we use a network with hidden layer of size $H = 100$, using  $L=5$ layers. The weights are initialized as outlined in \cite{sitzmann2020implicit}, i.e., by $[\bTheta_1]_{i,j} \propto \mathcal{U}\left(-\frac{1}{L},\frac{1}{L} \right)$ and  $[\bTheta_l]_{i,j} \propto \mathcal{U}\left(-\sqrt{\frac{c_0}{\omega_0^2H}},\sqrt{\frac{c_0}{\omega_0^2H}} \right)$ for $l=2,\hdots,L$, with $c_0=6$, which ensures that the distribution of the activation functions are preserved through the depth of the network. As in \cite{olivieri2024physics}, $\omega_0 = 30$.

\ninept

\bibliographystyle{IEEEbib}
\bibliography{export}

\end{document}